# User Interface (UI) Design Issues for Multilingual Users: A Case Study

Mahdi H. Miraz, Peter S. Excell, Maaruf Ali

*Abstract* -A multitude of web and desktop applications are now widely available in diverse human languages. This paper explores the design issues that are specifically relevant for multilingual users. It reports on the continued studies of Information System (IS) issues and users' behaviour across cross-cultural and transnational boundaries. Taking the BBC website as a model that is internationally recognised, usability tests were conducted to compare different versions of the website. The dependant variables derived from the questionnaire were analysed (via descriptive statistics) to elucidate the multilingual UI design issues. Using Principal Component Analysis (PCA), five de-correlated variables were identified which were then used for hypotheses tests. A modified version of Herzberg's Hygiene-motivational Theory about the Workplace was applied to assess the components used in the website. Overall, it was concluded that the English versions of the website gave superior usability results and this implies the need for deeper study of the problems in usability of the translated versions.

1. Introduction

THE motivation for this research was to examine in depth the UI Design issues for applications intended for bilingual and multilingual people. The case study reported has focused on the British Broadcasting Corporation's (BBC) website hosted at the URL www.bbc.co.uk. The reason for choosing this website for detailed study is that this is one of the most widely used global websites which caters for multiple language users.

The study was undertaken by using focus groups consisting of people from different countries, primarily including: Bangladesh, Egypt, India, Jordan, Pakistan and Saudi Arabia.

2. METHODOLOGY

Both qualitative and quantitative approaches were used to study the UI Design issues through the focus groups. The research consisted of direct observations of focus groups plus one-to-one tutoring sessions; structured and unstructured interviews were also included. To increase the reliability of the data, a multiple case-study approach was adopted and teams of tutors and researchers were employed for bias reduction [1,2] in the interpretation of the data. The Usability Factors were derived from the usability attributes and heuristics familiarized by Jakob Nielsen [3,4] the questionnaire suggested and used by him were also adopted and modified to meet the multilingual aspect of this research.

The reason for choosing the BBC website as the artefact for investigation is that the BBC claims that it is the global primary public service broadcaster. Its purpose is to "enrich people's lives with programmes that inform, educate and entertain" [5]. BBC News is the largest broadcast news gathering operation in the world and the BBC's online presence includes a comprehensive news website and archive [6]. According to Alexa's TrafficRank system [7], in July 2012, "BBC Online was the 27[th] most popular English Language website in the world and the 5[th] most popular in the UK and the 47[th] most popular overall". "Visitors to bbc.co.uk spend about seven minutes per visit to the site and 62 seconds per page-view. About 41% of the visitors to the site come from the UK and [the remaining] 59% of the visitors are from the rest of the world."

The focus groups, which consist of 164 voluntary participants, were given the task of exploring both the English and non-English versions of the BBC Online website (the non-English version being in the mother tongue of the users) they then provided their feedback on the design issues and usability of their website interaction. A survey [8] was also conducted using a questionnaire at the end of the study to verify the results. Participants were of both genders and it must be stressed that they were all volunteers aged between 18 to 80 years old. They were all from different cultural backgrounds and their first language was not English. The questionnaire was designed in bi-lingual format: English being the common language. The first languages of the participants were also used, especially: Arabic, Bangla, Hindi and Urdu.

Some of the questions were generated on the basis of detailed one-to-one interviews with the participants. This resulted in ten possible web components which were then subsequently used for analysing the hygiene-motivation factor part of the research.

M.H. Miraz, is a PhD Researcher at the Department of Computing, Glyndŵr University, UK and Lecturer at the Department of Computer Science & Software Engineering, University of Ha'il, KSA. Email: m.miraz@glyndwr.ac.uk or m.miraz@uoh.edu.sa

P.S. Excell, is a Professor of Communications at the Department of Computing as well as the Deputy Vice-Chancellor, Glyndŵr University, UK. Email: p.excell@glyndwr.ac.uk

M. Ali, is a Lecturer/Programme Leader in Computing in China, School of Architecture, Computing and Engineering, University of East London, UK. Email: maaruf@ieee.org

3. BACKGROUND

Although the World Wide Web is fundamental to communication in every sphere of life, even so Li and Kirkup [9] have said that 89% of websites are in the English language and primarily North American. Ishida [10] argued that the share of English web pages continues to decline whilst that of other languages continues to increase - it is thus paramount to ensure the multilingual success of the World Wide Web. So it comes as no surprise that the ability to understand the web page in one's own native language is one of the vital cross-cultural factors affecting attitudes towards the continued widespread adoption of Internet use.

By analogy with product design, Hermeking [11,12] stated that website design can in fact be said to be a specific set of economic, technical or instrumental, aesthetic and social qualities or symbolic attributes of a website that contribute to its users' satisfaction, which in turn relies on the users' values and cultural habits.

Recently there has been an increased perception of the need to design products and offer services for enhanced economic and social diversity. This is offered in the concept of 'design for all' principles or 'inclusive design' [13,14]. Designing for 'all' seeks to cater for wider usage [15]. Gesteland [16] states there are two "Great Divides" between business cultures: Deal Focus (DF) and Relationship Focus (RF). All the markets of the world except North America and Europe are in fact relationship-oriented, based on intricate personal networks. This business culture has a great impact on on-line shopping and e-business [17]. It is very probable that the language of the website is also an important factor here.

According to the WorldWatch Institute, around two-thirds of the world's population is classed as bilingual [18]. Naturally, there are thousands of people who learn a second language, either as children or adults, and there exist 250 languages in the world which have at least 1 million speakers each [18]. This is why a key reason for the success of the Internet is its ease of use and its cross-cultural and international reach. For clients who run global businesses, a way to complement cost effective international sales efforts with increased impact is by building a multilingual website platform [19].

Realising the importance of multilingual websites, the W3C organisation on 26-27th October, 2010, ran the first workshop in Madrid, entitled "The Multilingual Web: Where Are We?" hosted by the Universidad Politécnica de Madrid [10]. The aims were to survey and meet the challenges of the multilingual Web by introducing people to currently available standards and best practices which were aimed at helping content creators, tool developers, localizers and others. The primary objective was sharing information about the existing initiatives and identifying gaps.

Variations between cultures may give rise to many Web usability problems. These differences may be found in graphics, use of colours, icons, phrases, date and time format, character sets, symbols, pictures and so forth [20]. Different cultural users may comprehend the same websites in totally different ways. Misunderstanding, confusion and even offence to users may be caused by the inappropriate use of some metaphors, interaction sequences, appearance or navigation [21,22].

Miraz [23] et al. recently conducted a survey amongst the IS users of the UK and Bangladesh to determine how circumstances related to socio-economic situation and culture are being mirrored in the behaviour of IS users across different national boundaries and the relationship with the diffusion of mobile broadband technology (including Internet-based services) due to this. The study also outlined many issues affecting the IS users' behaviour, which included age and gender, education and economic capacity, language and so forth.

Elnahrawy [24] concludes that it is obligatory and not something optional to design websites that accommodate users from different languages and cultures. Online communications must address the language preferences of users. On the other hand, regardless of the language used, the online experience must be culturally relevant to the user in order to achieve an emotional connection (or engagement) with the intended audience.

4. DESIGN ISSUES

A multilingual website offers content in more than one language whereas a multi-regional website is unequivocally aimed for users in different countries. Creating an individual experience for a site owner's consumers, regardless of the language they speak or where they are based, or scaling a website to cover multiple languages and/or countries, can be an enormously complicated task, not least in synchronisation of updates. In this section, the UI design issues relating to multilingual websites will be discussed and the data collected through the survey will be analyzed. The focus will be on the issues discussed in the following.

*4.1. Language Selection and Usability*

Automated selection of language might be sometimes frustrating, especially for immigrant website visitors. For example, Skype.com automatically detects the IP (Internet Protocol) address and diverts the user accordingly. However, if a non-Arabic speaker resident of Saudi Arabia wants to visit the Skype website, because it is automatically directed to the Arabic version and the option to navigate to the English/other version is presented in Arabic, there is no easy way of doing that for anyone unable to read Arabic. Changing language is often offered graphically by national flags, but this poses a risk of controversy. For example, English, French, Arabic, Bangla, Spanish and so forth languages are spoken by more than one nationality. Introducing one unique symbol of each language could solve the problem but due to localized variations of such languages may lead to further difficulties.

Such scenarios may be circumvented by allowing the visitors to choose their region and then language upon initial interaction with the website application. This can be really effective for websites that are going global. Being able to choose Saudi Arabia and then select Bangla or English, for example, can make a brand seem niche and unique for the visitor. Thus, local content can be provided in a language of the user's own choice. Usability of the website can be increased by making it even easier for the visitors. IP can be tracked to select the region and then auto-detection of the browser language could take place so that the website could be automatically served in that language. A 'change back' option should always be present to facilitate the visitors in case the users want to visit pages of some other regions or languages or even if they relocate (periodically).

*4.2. Graphics and Placement of Text and Images*

People from different languages and cultures read in different ways. For instance, Semitic cultures read from right-to-left, while most Western cultures read from left-to-right and Pacific-Oceanic cultures read vertically from top-to-bottom in columnar format. A navigation bar may be totally unsuitable on the right for one culture but may be perfectly normal for another. Thus placement of images and text can have a vital role for the overall usability, conversion and acceptance of a site. 51.2% of participants consider the graphics displayed in the English version of the BBC website to be clear and attractive and 37.8% consider it to be satisfactory to some extent, as shown in Table 1. On the other hand, more participants are found to be unhappy with the non-English version, as shown in Table 1. Only 35.4% of the participants consider the graphics of the non-English version to be clear and attractive and 51.2% consider them to be satisfactory to some extent. One interesting finding is that 53.7% of the participants consider that the graphics presented in the English version add to readability of the site and 15.9% consider them as detracting from readability. In contrast, 56.7% of the participants consider that the graphics presented in the non-English version add to readability of the site and 18.3% consider them as detracting from readability. The Hypothesis test for the PCA-derived Textual Graphics component also did not provide any significant result (see Section: 6.4). As these results do not lead to any firm conclusions, contrary to expectations, further research is needed to clarify this issue.

Table 1. Usefulness of Website's Graphics: Responses

| Usefulness of Graphics clarity and attractiveness | English Version | | Non-English Version | |
|---|---|---|---|---|
| | No. of Responses | Percentage | No. of Responses | Percentage |
| Yes | 84 | 51.2% | 58 | 35.4% |
| To some extent | 62 | 37.8% | 84 | 51.2% |
| No | 18 | 11.0% | 22 | 13.4% |
| Total observation | 164 | 100% | 164 | 100% |

*4.3. Colour*

Colour has a range of culturally dependent significances and thus it is very important to take this into consideration when designing multilingual and multiregional websites. In the Western World for example, red is often the colour of love while in South Africa it denotes the colour of misery and mourning. Testing has in fact shown that efficiency in the intelligibility and the interactivity rate increased in Russia when using a black and red combination for a "Call to Action" button whilst in Italy this was maximised by changing it from red to orange [25]. 39.6% of participants considered the colour of the English version to be attractive and appealing, 54.3% considered the colour to be attractive to some extent and the remaining 6.1% were of the opinion that the colours were not attractive, as shown in Figure 1. On the other hand, the number of satisfied participants for the non-English version was much lower. Only 29.3% of participants considered the colour of the non-English version to be attractive and appealing, 48.2% considered the colour to be attractive to some extent and the remaining 22.6% were of the opinion that they were not attractive, as shown in Figure 2. This clearly indicates that more care should be taken while developing a website for a wide diversity of visitors when utilising colour in the web page content.

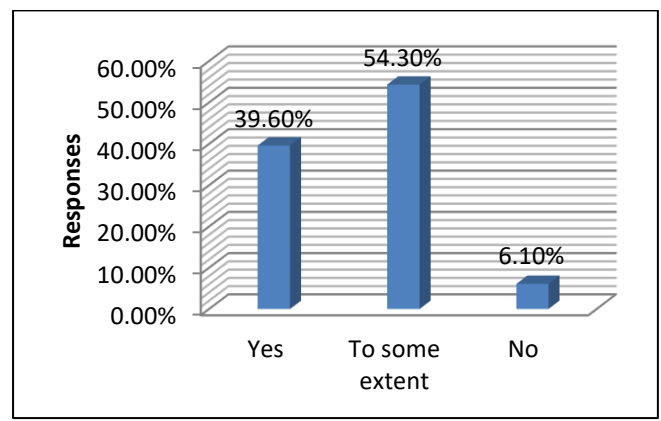

Fig. 1. English Website Attractiveness and Appeal of Colours.

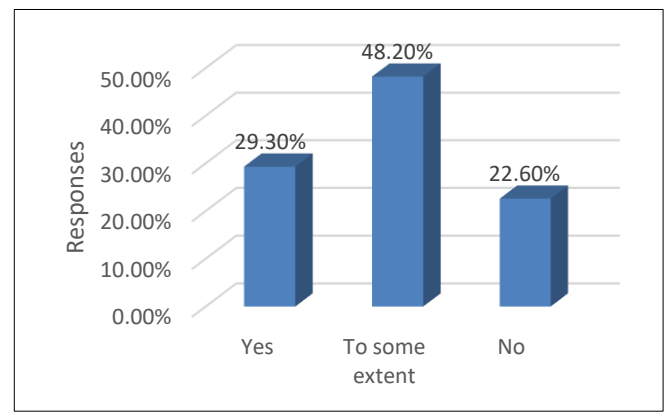

Fig. 2. Non-English Website Attractiveness and Appeal of Colours.

### 4.4. Translation

There exist many automated translation facilities that can be used on websites, some even free. But automated translation does not really help. In most of the cases, such facilities failed to translate the focused meaning, especially when metaphors, cultural terms or terms having multiple meanings were involved. Even extra care is needed while translating the contents manually. If demand justifies it, expert translators from appropriate cultures/origins should be included in the design and development team. Cap [26] conducted a study on the neutrality of Wikipedia (WP), as shown in Table 2, by analysing how pages dated March 2010 covering a highly controversial figure were reported based on the website's geographical origin and using the example of "Bin Laden".

Table 2. Translation and Description of Bin Laden.

| Language | Reporting Perception |
|---|---|
| **German** | as a terrorist |
| **English** | a leader of a terrorist organization |
| **Arabic** | "founder and leader of al-Qaeda network" |
| **Hebrew** | "terrorist leader of the Islamic terrorist organization Al-Qaeda" |
| **Chinese** | "leader [of] the organization [...] a lot of people think that is a global terrorist organization". |

Thus deviations in the neutrality of websites can clearly be seen, based on the geographical region where the website is compiled.

The present survey found that only 23.8% of the participants considered that the translation of the Non-English version matched closely to the original news in the English version. 61.6% were of the opinion that it matched to some extent and the remaining 14.6% were not totally satisfied with the translation. This is shown in Figure 3, below.

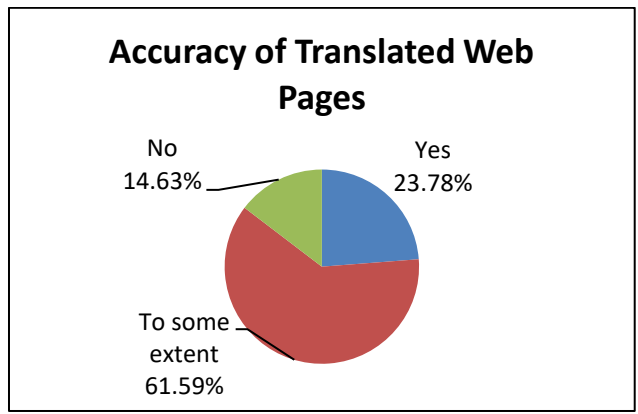

Fig. 3. Perception of the Accuracy of Translated Web Pages.

### 4.5. Abbreviations and Keywords

Some abbreviations, such as FAQ, ASAP are very commonly used in the English language. Finding their non-English version/translation is sometimes very difficult. Extra care should be taken while translating any abbreviations because in most cases suitable alternatives cannot be found or simply do not exist. The survey showed that only 27.4% of participants considered that the abbreviated words had suitable alternatives in the Non-English versions, 40.2% considered them to be suitable to some extent and the remaining 32.3% considered them unsuitable. This is illustrated in Figure 4.

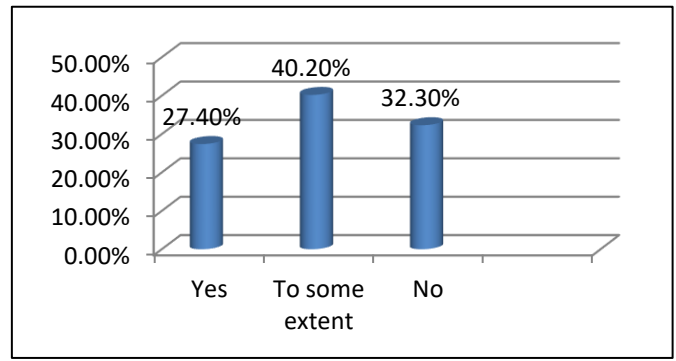

Fig. 4. Existence of Abbreviations in Multiple Languages.

As with translations and abbreviations, keywords do not "literally" correlate either. For instance, 'Cheap Flights' in the US gets 6.12 million searches per month and in Italy '*voli economici*', the literal translation, gets [only] 33,000 searches per month. If that is tweaked for the culturally meaningful '*voli* low cost', the rate rises to 246,000 searches per month. That is a significantly large difference [25].

### 4.6. Localizing the website

Translating the website into a local language is not the only aspect of localization. Consideration should also include: weather, weights and measures format, currency symbols and conversions, date format, government holidays, cultural sensitivities, geographic examples and gender roles, beliefs and religions, traditions and social structures, level of fluency with ICT, education and so forth. The survey revealed that 42.1% of the participants agreed that the news on the non-English versions is localized. Among them, more than half (59.8%) would like the news to be localized. The results are displayed in Figure 5. More on this issue is discussed under the "Motivation Theory" section of this paper.

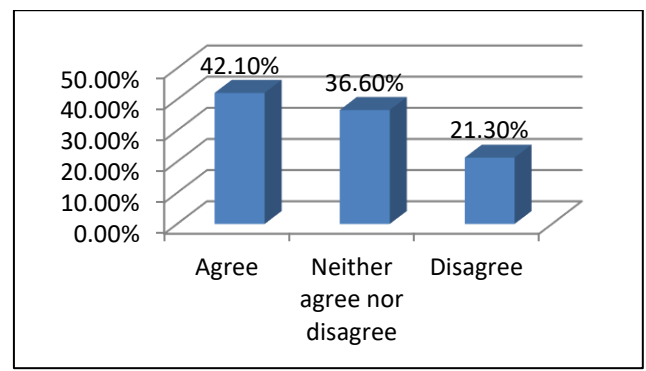

Fig. 5. Language Localization of the News.

### 4.7. Page layout and Navigation

The survey found that only 1.2% of the participants considered the layout of the English version to be not easy to understand, as shown in Figure 6.

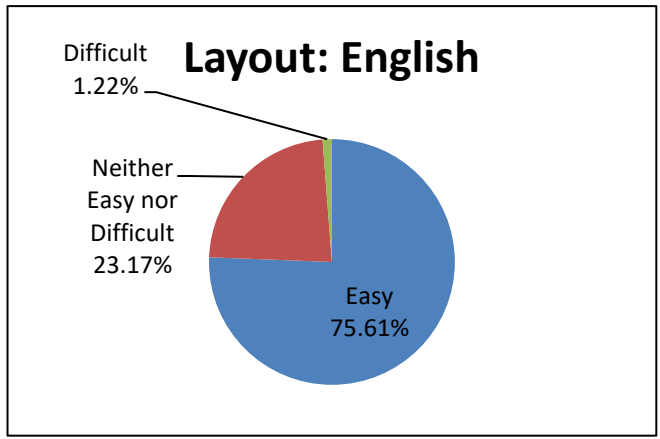

Fig. 6. Web Layout – English Version.

This number however was higher (8.5%) for the non-English version of the site, as shown in Figure 7. This issue can be solved to an increased extent by adopting the same design techniques. When converting from a left-to-right script to right-to-left, a simple approach may be to adopt a mirror image layout translation of the website. Some images may also need to be sequenced in the right order when showing events.

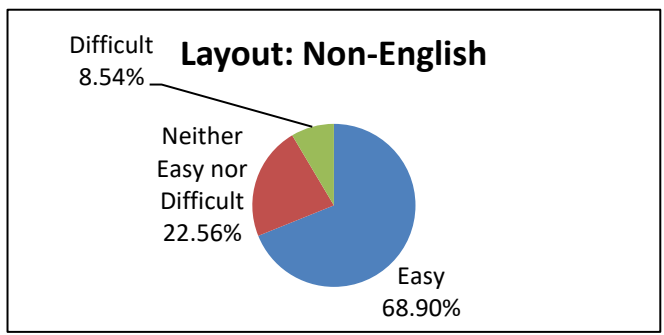

Fig. 7. Web Layout – Non-English Version.

Figure 8 presents the respondents' results for the navigability of the BBC English website and Figure 9 illustrates the navigability of the BBC non-English website. The results show a marginal preference of the non-English website in terms of page navigability of 6.1% over the English version. This contrasts with the page layout preference of 6.7% of the English version over the non-English version. Thus, further study is required to draw more statistically significant results for the navigability and layout issues for the English and non-English versions of the web sites.

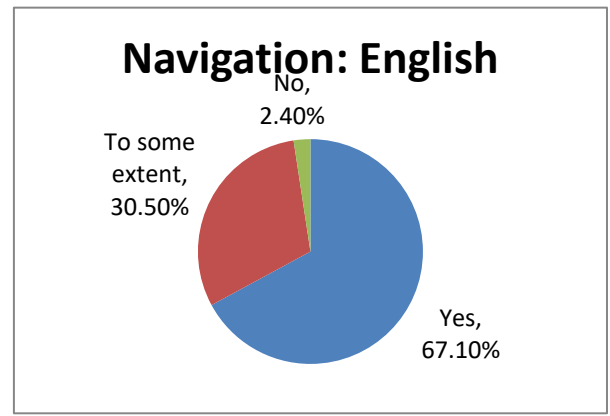

Fig. 8. Navigability of English BBC Website.

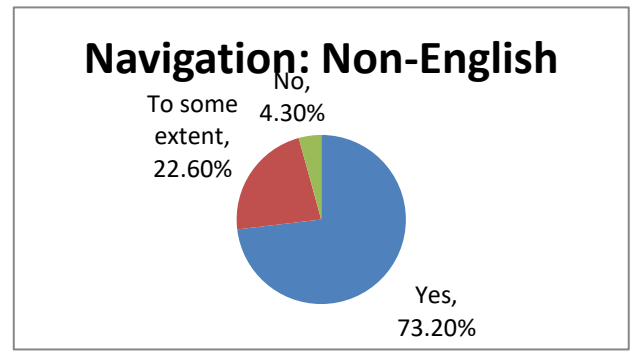

Fig. 9. Navigability of Non-English BBC Website.

### 4.8. Font Size and Legibility of Websites

Some oriental languages, such as Arabic, Chinese, and Korean are difficult to read at font sizes that are perfectly legible for European languages like English, French and Russian. The survey reveals that 67.7% of participants consider the fonts of the English version to be readable, attractive and properly sized, 26.8% consider them to be adequate to some extent and 5.5% are of the opposite opinion, as shown in Figure 10. In contrast, 61.6% of participants consider the fonts of the non-English version to be readable, attractive and properly sized, 26.2% consider them adequate to some extent and 12.2% are of the opposite opinion, as shown in Figure 11. The result clearly indicates that more attention is required when choosing fonts for a multi-lingual website.

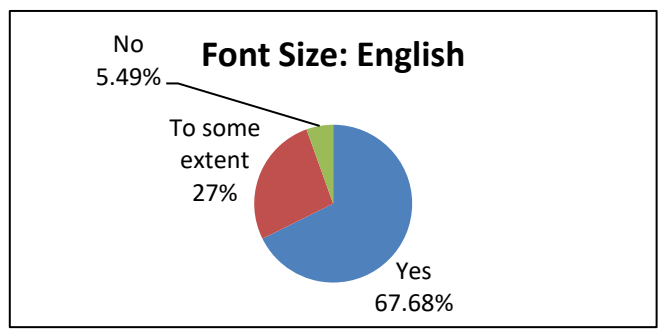

Fig. 10. Font Size Legibility of English Website.

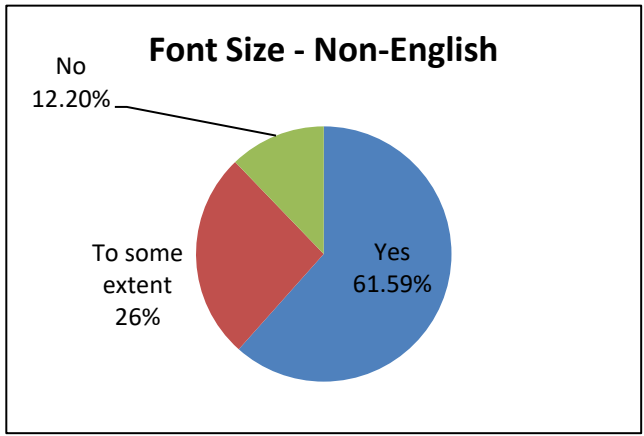

Fig. 11. Font Size Legibility of Non-English Website.

*4.9. Fitting the Text into Web Pages*

Text in some languages may take up more space than others; for example, German and Russian compared to English, due to their expressive nature; whilst Chinese and Korean take less space. Certain Web sections, especially menus, often have fixed widths and it then becomes necessary to use a shorter translation to fit the available space [27]. The survey has found that 64.6% of the participants find no difficulty in reading the English version but this percentage is lower (55.5%) for the non-English versions. For the non-English version, 16.5% of participants find difficulty reading the texts and another 28.1% find it difficult to some extent. These results are shown in Figures 12 and 13.

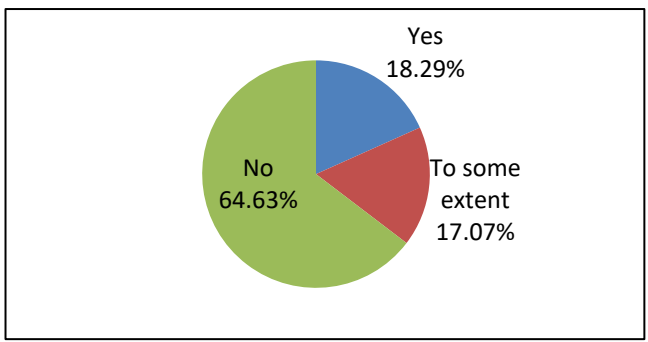

Fig. 12. Difficulty Reading Texts: English.

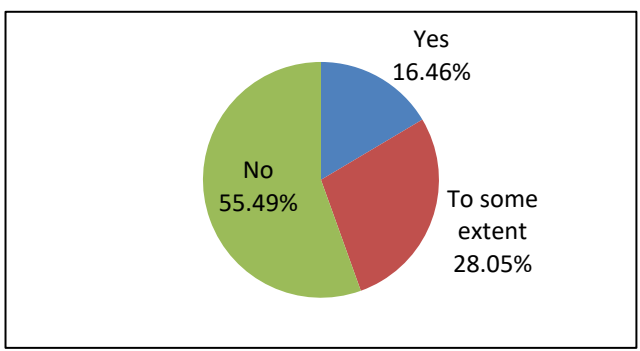

Fig.13. Difficulty Reading Texts: Non-English.

5. DATA ANALYSIS (USING SPSS)

Data were collected using dependent variables to get the user responses regarding the two different versions of the website they explored (English and Non-English versions). Factor Analysis (FA) using Principal Component Analysis (PCA) was then conducted to de-correlate the variables and thus reduce their numbers. Two different FA were run, based on the fact that the data were collected from a dependent sample under two different experimental conditions (i.e. versions of the website). However, the components generated by the two different FA using SPSS were not unique. So a further FA was conducted, considering the data to be from an independent sample.

The reliability of FA is also dependent on the sample size. Field [28] reviews many suggestions about the sample size necessary for reliable factor analysis and concludes that it depends on many things. In general over 300 cases are probably adequate but communalities after extraction should probably be above 0.5. The data of this study satisfied both of the conditions and hence FA was proceeded with.

Tabachnick and Fidell [29] concluded that the Anderson–Rubin method is best utilised when uncorrelated scores are required though the regression method is preferred in other circumstances simply because it is the most easily understood. Hence, the Anderson–Rubin method was chosen for the analysis. Further, Varimax Orthogonal Rotation was applied [28] because it is an elegant approach that simplifies the interpretation of factors. The option 'Suppress Absolute Values' was set to be less than 0.4: this ensures that factor loadings within ±0.4 are not displayed in the output.

Scanning the Correlation Matrix table produced by SPSS did not reveal any variable for which the majority of significance values were greater than 0.05. No correlation coefficient values greater than 0.9 could be found after scanning the correlation coefficient table. As a result, it can be assured that no problem could arise because of singularity in the data.

The determinant for these data is 0.061, which is greater than the necessary value of 0.00001. Therefore, the existence of multicollinearity is not a problem for these data.

Table 3 shows the result of the Kaiser-Meyer-Olkin (KMO) measure of sampling adequacy and Bartlett's test of Sphericity. Kaiser [30] recommends accepting values greater than 0.5 as acceptable. Bartlett's test of sphericity produces $\chi^2$ (105) = 898.028, $p < 0.001$ and the value of KMO in this case is 0.732, which is good according to Hutcheson and Sofroniou [31] and Field [28]. KMO is greater than 0.5, so the sample was adequate for FA.

Table 3. KMO and Bartlett's Test.

| Kaiser-Meyer-Olkin Measure of Sampling Adequacy. | | 0.732 |
|---|---|---|
| Bartlett's Test of Sphericity | Approx. Chi-Square | 898.028 |
| | df | 105 |
| | Sig. | 0.000 |

Table 4. Rotated Component Matrix[a]

| | Component | | | | |
|---|---|---|---|---|---|
| | 1 | 2 | 3 | 4 | 5 |
| Are graphics clear and attractive? | 0.751 | | | | |
| Are elements sized and arranged to fit within browser window? | 0.751 | | | | |
| Are colours attractive and appealing to you? | 0.552 | | | | |
| Is there a good use of "white space"? | 0.498 | | 0.450 | | |
| Can you navigate readily from page to page? | | 0.814 | | | |
| Is the site layout easy to understand? | | 0.678 | | | |
| Are the pages easy for the visitor to read? | 0.477 | 0.507 | | | |
| Is it easy to get back to Home page or top of a page? | | 0.474 | | | |
| Do you find difficulty reading any text(s) due to colour combination? | | | 0.768 | | |
| Are the types of fonts readable, attractive and properly sized? | | 0.488 | -0.491 | | |
| Do the graphics (add to/detract from/neither) readability? | | | | 0.776 | |
| Do the page elements follow a logical sequence? | | | | 0.581 | 0.408 |
| Do graphics contribute to the purpose of the page? | | | | 0.532 | |
| Do the links look the same in different browsers and are easy for the visitor to spot? | | | | | 0.715 |
| Do the pages look good with various browsers? | | | | | 0.671 |

Extraction Method: Principal Component Analysis.
Rotation Method: Varimax with Kaiser Normalization.
a. Rotation converged in 8 iterations.

Taking all of these statistical conditions into account, all of the designed questions used in the survey correlated adequately and none of the correlation coefficients were excessive; therefore, there was no need to consider eliminating any questions at this stage.

By Kaiser's criterion, four factors should be extracted. However, this criterion is only accurate when:
(a) there are fewer than 30 variables and the communalities after extraction are greater than 0.7, or
(b) the sample size exceeds 250 and the average communality is greater than 0.6.

None of the communalities calculated by SPSS for these data exceed 0.7. The average of the communalities is 8.662/15 = 0.578. So, on both grounds adopting Kaiser's rule may not be accurate in the present case. However, the literature on Kaiser's criterion gives recommendations for much smaller samples. It is also possible to use the Scree Plot as an output from SPSS, as shown in Figure 14. This curve is difficult to interpret because it begins to tail off after three factors, however there is another drop after five factors before a stable plateau is reached. Therefore, it appears reasonable to retain either two or five factors. At this stage, for these data SPSS had extracted five factors and hence it was concluded to be safe to retain these five factors for the final analysis.

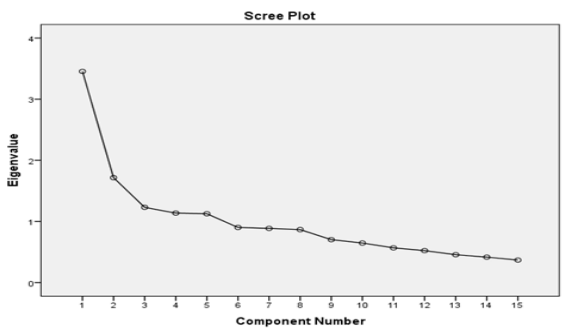

Figure 14. Scree Plot of the Component Loading.

Table 4 shows the SPSS output of the Rotated Component Matrix. The items that cluster around the same components suggest that Component 1 represents variables related to Textual Graphics, Component 2 is related to Layout and Navigation, Component 3 for Readability, Component 4 about Non-textual Related Graphics and Component 5 for matters related to Cross-browser Compatibility. These components were saved as variables and then used for further analysis.

These PCA-derived variables were used for the hypothesis tests. With a sufficiently large population of sample sizes (>30 or 40), the normality assumption should not be in violation and thus no major problems should arise [32]; thus the implication is that parametric procedures may be used even when the data are not normally distributed [33]. So samples consisting of hundreds of observations indicate that the data distribution can be ignored [34]. In keeping with the Central Limit Theorem, the following can also be stated:
(a) if the sample data are approximately normal then the sampling distribution too will be normal;
(b) in large samples (>30 or 40), the sampling distribution tends to be normal, regardless of the shape of the data [28] [33]; and
(c) means of random samples from any distribution will themselves have normal distribution [34].

As a result, the Independent-means $t$-test was used to test hypotheses.

## 6. HYPOTHESES TESTING

Based on the components derived by PCA, the following hypotheses were constructed and tested to investigate the usability and design issues of multilingual websites in further details.

## A. Non-Textual Graphics

For Non-Textual Graphics the following alternative and null hypotheses were considered:

Ha = The usages of Non-Textual Graphics in the English version of webpages have superior intelligibility over the Non-English version.

Ho = The usages of Non-Textual Graphics in the English version of webpages have no effect on the intelligibility over the Non-English version.

For Levene's Test (As shown in Table 5) for the Equality of Variances, $F = 0.948$, Sig. $= 0.331 > 0.05$. Since at the 95% confidence interval level, the variances are not significantly different, the results of the top row of the t-test (as shown in Table 6) should be utilised.

Table 5. Group Statistics (Non-Textual Graphics).

|  | Version of the Website | N | Mean | Std. Deviation | Std. Error Mean |
|---|---|---|---|---|---|
| FA_Non_Textual_Graphics | English | 164 | -0.1649 | 0.9941 | 0.0776 |
|  | Non-English | 164 | 0.1649 | 0.9814 | 0.0766 |

Assuming equal variances, for the *t*-test, $t = -3.023$ with 326 degrees of freedom, Sig. $= 0.0015 < 0.05$ (for 1-tailed). Hence $H_0$ was rejected at the 5% significance level.

The sample evidence indicates that the usage of Non-Textual Graphics in the English version of webpages has superior intelligibility over the Non-English version.

Table 6. Independent Samples Test (Non-Textual Graphics)

| | | Levene's Test for Equality of Variances | | *t*-test for Equality of Means | | | | | | |
|---|---|---|---|---|---|---|---|---|---|---|
| | | F | Sig. | t | df | Sig. (2-tailed) | Mean Difference | Std. Error Difference | 95% Confidence Interval of the Difference | |
| | | | | | | | | | Lower | Upper |
| FA_Non_Textual_Graphics | Equal variances assumed | 0.948 | 0.331 | -3.023 | 326 | 0.003 | -0.3297 | 0.10908 | -0.5443 | -0.1151 |
| | Equal variances not assumed | | | -3.023 | 326 | 0.003 | -0.3297 | 0.10908 | -0.5443 | -0.1151 |

## B. Layout and Navigation

For Layout and Navigation the following alternative and null hypotheses were considered:

Ha = The Layout and Navigability in the English version of webpages is easier to comprehend and steer compared to the Non-English version.

Ho = The Layout and Navigability in the English version of webpages has no effect on the comprehension and steerability compared to the Non-English version.

For Levene's Test (As shown in Table 7) for the Equality of Variances, $F = 25.938$, Sig. $= 0.000 < 0.05$. Since at the 95% confidence interval level, the variances are significantly different, the results of the bottom row of the t-test (as shown in Table 8) should be utilized.

Table 7. Group Statistics (Layout and Navigation).

|  | Version of the Website | N | Mean | Std. Deviation | Std. Error Mean |
|---|---|---|---|---|---|
| FA_Layout_and_Navigation | English | 164 | -0.1107 | 0.8238 | 0.0643 |
|  | Non-English | 164 | 0.1107 | 1.1414 | 0.0891 |

Not assuming equal variances, for the *t*-test, $t = -2.014$ with 296.6 degrees of freedom.

Sig. $= 0.0225 < 0.05$ (for 1-tailed). $H_0$ is rejected at the 5% significance level.

The sample evidence indicates that the layout and navigability in the English version of webpages is easier to comprehend and steer compared to the Non-English version.

Table 8. Independent Samples Test (Layout and Navigation).

| | | Levene's Test for Equality of Variances | | *t*-test for Equality of Means | | | | | | |
|---|---|---|---|---|---|---|---|---|---|---|
| | | F | Sig. | t | df | Sig. (2-tailed) | Mean Difference | Std. Error Difference | 95% Confidence Interval of the Difference | |
| | | | | | | | | | Lower | Upper |
| FA_Layout_and_Navigation | Equal variances assumed | 25.938 | .000 | -2.01 | 326 | .045 | -0.2214 | 0.10992 | -0.4377 | -0.0052 |
| | Equal variances not assumed | | | -2.01 | 296.6 | .045 | -0.2214 | 0.10992 | -0.4377 | -0.0051 |

## C. Readability

For Non-Textual Graphics the following alternative and null hypotheses were considered:

Ha = The English version of webpages has enhanced readability over the Non-English version.

Ho = The English version of webpages has no difference in readability over the Non-English version.

Table 9. Group Statistics (Non-Textual Graphics).

| | Version of the Website | N | Mean | Std. Deviation | Std. Error Mean |
|---|---|---|---|---|---|
| FA_Readability | English | 164 | 0.0526 | 1.0730 | 0.0838 |
| | Non-English | 164 | -0.0526 | 0.9216 | 0.0720 |

For Levene's Test (as shown in Table 9) for the Equality of Variances, F = 1.251, Sig. = 0.264 > 0.05. Since at the 95% confidence interval level, the variances are not significantly different, the results of the top row of the t-test (as shown in Table 10) should be utilized.

Table 10. Independent Samples Test (Non-Textual Graphics).

| | | Levene's Test for Equality of Variances | | t-test for Equality of Means | | | | | | |
|---|---|---|---|---|---|---|---|---|---|---|
| | | F | Sig. | t | df | Sig. (2-tailed) | Mean Difference | Std. Error Difference | 95% Confidence Interval of the Difference | |
| | | | | | | | | | Lower | Upper |
| FA_Read-ability | Equal variances assumed | 1.251 | 0.264 | 0.952 | 326 | 0.342 | 0.10512 | 0.11045 | -0.11216 | 0.32240 |
| | Equal variances not assumed | | | 0.952 | 319 | 0.342 | 0.10512 | 0.11045 | -0.11218 | 0.32242 |

Assuming equal variances, for the *t*-test, *t* = 0.952 with 326 degrees of freedom,

Sig. = 0.171 > 0.05 (for 1-tailed). Hence, H₀ at the 5% significance level has been accepted.

The sample evidence indicates that the English version of webpages has no difference in readability over the Non-English version.

## D. Textual Graphics

For Textual Graphics the following alternative and null hypotheses were considered:

Ha = The usages of Textual Graphics in the English version of webpages has superior intelligibility over the Non-English version.

Ho = The usages of Textual Graphics in the English version of webpages has no effect on the intelligibility over the Non-English version.

For Levene's Test (as shown in Table 11) for the Equality of Variances, F = 1.202, Sig. = 0.274 > 0.05. Since at the 95% confidence interval level, the variances are not significantly different, the results of the top row of the *t*-test (as shown in Table 12) should be utilized.

Table-11 Group Statistics (Textual Graphics)

| | Version of the Website | N | Mean | Std. Deviation | Std. Error Mean |
|---|---|---|---|---|---|
| FA_Textual_Graphics | English | 164 | 0.0858 | 0.9802 | 0.0765 |
| | Non-English | 164 | -0.0858 | 1.0152 | 0.0793 |

Table-12 Independent Samples Test (Textual Graphics)

| | | Levene's Test for Equality of Variances | | t-test for Equality of Means | | | | | | |
|---|---|---|---|---|---|---|---|---|---|---|
| | | F | Sig. | t | df | Sig. (2-tailed) | Mean Difference | Std. Error Difference | 95% Confidence Interval of the Difference | |
| | | | | | | | | | Lower | Upper |
| FA_Textual_Graphics | Equal variances assumed | 1.202 | 0.274 | 1.557 | 326 | 0.121 | 0.17152 | 0.11019 | -0.04526 | 0.38830 |
| | Equal variances not assumed | | | 1.557 | 326 | 0.121 | 0.17152 | 0.11019 | -0.04526 | 0.38830 |

Assuming equal variances, for the *t*-test, *t* = 1.557 with 326 degrees of freedom,

Sig. = 0.0605 > 0.05 (for 1-tailed). Hence, H₀ at the 5% significance level has been accepted.

The sample evidence indicates that the usages of Textual Graphics in the English version of webpages have no effect on the intelligibility over the Non-English version.

## E. Cross-browser Compatibility

For Non-Textual Graphics the following alternative and null hypotheses were considered:

Ha = The Cross-browser Compatibility of the English version of webpages has more uniform behaviour compared to the Non-English version.

Ho = The Cross-browser Compatibility of the English version of webpages has no significant difference in behaviour compared to the Non-English version.

For Levene's Test (as shown in Table 13) for the Equality of Variances, F = 7.786, Sig. = 0.006 < 0.05. Since at the 95% confidence interval level, the variances are significantly different, the results of the bottom row of the *t*-test (as shown in Table 14) should be utilised.

Table 13. Group Statistics (Cross-browser Compatibility).

| | Version of the Website | N | Mean | Std. Deviation | Std. Error Mean |
|---|---|---|---|---|---|
| FA_Cross_Browser_Compatibility | English | 164 | -0.0376 | 0.9450 | 0.0738 |
| | Non-English | 164 | 0.0376 | 1.0537 | 0.0823 |

Not assuming equal variances, for the *t*-test, *t* = -0.680 with 322.210 degrees of freedom, Sig. = 0.2485 > 0.05 (for 1-tailed). $H_0$ at the 5% significance level has been accepted.

Table 14. Independent Samples Test (Cross-browser Compatibility).

| | | Levene's Test for Equality of Variances | | *t*-test for Equality of Means | | | | | | |
|---|---|---|---|---|---|---|---|---|---|---|
| | | F | Sig. | t | df | Sig.(2tailed) | Mean Difference | Std. Error Difference | 95% Confidence Interval of the Difference | |
| | | | | | | | | | Lower | Upper |
| FA_Cross_Browser_Compatibility | Equal variances assumed | 7.786 | 0.006 | -0.680 | 326 | 0.497 | -0.075194 | 0.1105223 | -0.2926 | 0.14223 |
| | Equal variances not assumed | | | -0.680 | 322.2 | 0.497 | -0.075194 | 0.1105223 | -0.2926 | 0.14224 |

The sample evidence indicates that the Cross-browser Compatibility of the English version of webpages has no significant difference in behaviour compared to the Non-English version.

## 7. APPLICATION OF THE MODIFIED HERZBERG'S HYGIENE-MOTIVATIONAL THEORY

Zhang [35,36] argued that Herzberg's Hygiene-motivational Theory about the Workplace [37,38] could be utilised in the Web environment. The hygiene-motivational theory, also known as the dual-factor theory, was introduced by psychologist Frederick Herzberg [39] back in 1959. The theory positions that certain factors in the workplace cause job satisfaction, while a different set of factors cause dissatisfaction. Thus it was theorized that job satisfaction and job dissatisfaction act independently of each other.

By using this analogy for website design, the presence of such hygiene factors would provide the basic functionality of a website, while their absence would create dissatisfaction in the users. Motivating factors are those that contribute to user satisfaction. They add additional value and may attract users back to that website. The authors of this paper are of the opinion that the usage of this theory can be further extended for usability testing and enhancement of other IS products as well.

While considering inclusion of the hygiene or motivating factors to be examined, web components present in the English version but absent in the Non-English versions and vice-versa have been considered. In addition, in Herzberg's Original Hygiene-motivational method about Workplace research, only binary (yes/no) responses were used to classify the variables either as hygiene factor or motivational factor. However, some participants might be neutral and in fact be driven by both. Hence in the questionnaire used here, the option to be neutral was included. For descriptive statistical analysis, a "Yes" response was ranked as 1; "It doesn't bother me" as 0 and "No" as -1. Based on the responses provided, any components having the mean value of 0.5 or more are considered as the hygiene factors and values less than 0.5 are considered as the Motivating factors for most visitors.

Firstly, the web components which are available for the English version but not present in the Non-English version of the site were considered. The components taken into consideration were: 'Share Market Information', 'Weather', 'Health Information', 'Magazine', 'Sponsored Advertisements', 'Learning English', 'Watching News Summary' and 'Top News Story'. Figure 15 presents the responses in percentage for Non-English website.

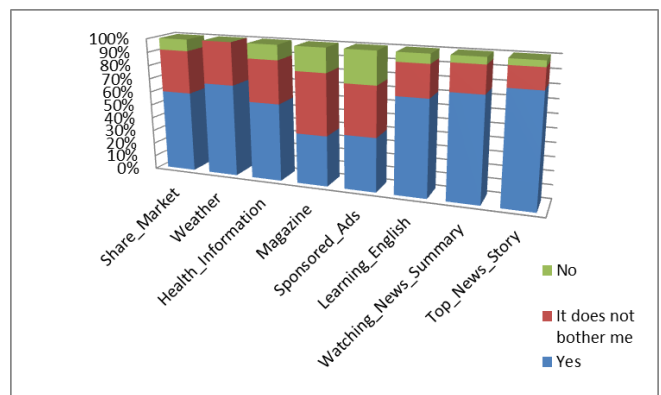

Fig. 15. Responses in Percentage for Non-English Website.

Table 15 presents the descriptive statistical analyses (Mean and Standard Deviation) of the collected responses.

Table 15. Descriptive Statistics of the Responses for Non-English Version of the Website

| Factor | Mean value | Std. deviation |
|---|---|---|
| Share_Market | 0.51 | 0.65 |
| Weather | 0.68 | 0.47 |
| Health_Information | 0.46 | 0.69 |
| Magazine | 0.20 | 0.72 |
| Sponsored_Adds | 0.16 | 0.78 |
| Learning_English | 0.63 | 0.61 |
| Watching_News_Summary | 0.71 | 0.55 |
| Top_News_Story | 0.77 | 0.52 |

According to the threshold set, 'Share Market Information', 'Weather', 'Learning English', 'Watching News Summary' and 'Top News Story' have been identified as Hygiene factors. 'Health Information', 'Magazine' and 'Sponsored Advertisements' have been identified as Motivating Factors.

Next, the web components which are available for the Non-English version but not present in the English version of the site were considered. Based on the interviews conducted, only two components for possible inclusion in this experiment could be identified. The components considered were: 'Social Networking' and 'Site Survey'. Figure 16 presents the Responses in Percentage for English Website.

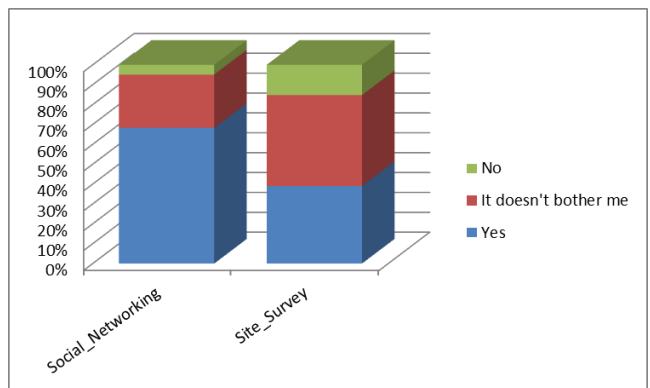

Fig. 16. Responses in Percentage for English Website.

Table 16 shows the descriptive statistical analyses (Mean and Standard Deviation) of the collected responses.

Table 16. Descriptive Statistics of the Responses for English Version of the Website

| Factor | Mean value | Std. deviation |
|---|---|---|
| Social_Networking | 0.63 | 0.58 |
| Site_Survey | 0.24 | 0.70 |

According to the threshold set in this study, the Social Networking tool has been identified as a Hygiene factor whereas Site Survey has been identified as a Motivating Factor.

8. CONCLUSION

A subjective analysis based on a detailed questionnaire of users of the English and the non-English versions of a major international website found a divergence in the usability between them. The participants were specifically chosen to be users of English as a second language: they reported that the English sites were superior overall in terms of various parameters such as design and interactivity.

An extensive case-study was conducted across various geographical and linguistic groups, the scope of which the authors could not find in their literature survey to asses variations in the user experience of the English and non-English versions of websites. The research also addressed the key differences between multilingual and multiregional websites. The questionnaire based analyses did in fact show divergence in the interactivity and user satisfaction. As far as is known by the authors, this novel multilingual aspect of the research has not been addressed in the relevant literature.

An often overlooked usability factor regarding the language selection feature in multilingual or multiregional websites is that it is predominantly offered in a language unknown to the users. This frequently causes the users to completely abandon further interactivity with the website because they are unable to easily select the language they understand. The research concluded that combining the IP address as well as the language of the users' browser would eliminate this problem.

The survey found that the careful juxtaposition of images and text can increase the overall usability, conversion and acceptance of a site. The use of colour in designing a multilingual website has a significant impact that is often overlooked by the web designers. The case study presented found that colour preference varies among different socio-cultural groups and this had a direct impact on the speed of interactivity and overall satisfaction.

Automated translation does not address the many issues such as abbreviations, metaphors and cultural terms. Thus, trained human translators are still necessary to meet the visitor's level of expectations and satisfaction. Translation also has to encompass converting such factors as measurement units, date formats, currency symbols, gender roles, beliefs and religions so forth.

The survey also concluded that pictographic and ideographic - based characters need to be rendered often at a larger size than textual based characters. This indicates that space has to be reserved to accommodate these characters while designing the multilingual websites. The participants prefer the layout and navigability of the English version of the website. These indicate that more planning and attention need to be given while designing and developing the non-English version to provide the similar level of satisfaction.

Using PCA, the variables derived from the user questionnaire were de-correlated to just five. They were then used for the hypothesis tests. These confirmed the results found by using the subjective questionnaire, namely that the use of non-textual graphics, layout and navigability were found to be superior in the English version of webpages. However no significant differences in usability were found for textual graphics, readability and cross-browser compatibility.

Using an extended and modified version of Herzberg's Hygiene-Motivational Theory, the research identified ten components to be considered as either hygiene-oriented or motivational for multilingual websites. The research concluded that for the English version of the website: 'Social Networking' was found to be a Hygiene factor whereas 'Site Survey' was found to be a Motivational factor. For non-English counterparts: 'Share Market Information', 'Weather', 'Learning English', 'Watching News Summary' and 'Top News Story' have been identified as Hygiene factors, whereas 'Health Information', 'Magazine' and 'Sponsored Advertisements' have been identified as Motivating Factors.

The findings indicate that non-English versions of web pages need to be improved if they are to be fully equivalent to their original English version. The findings also concluded that there is scope for further analysis and usability design improvements in the use of graphics in web pages for both English and non-English versions of such websites.